\documentclass[twocolumn,prl,aps,showpacs]{revtex4}
\usepackage{amssymb}

\usepackage{amsmath}
\usepackage{graphicx}
\usepackage{dcolumn}
\usepackage{bm}

\begin{document}

\preprint{APS/123-QED}
\title{Formation Mechanism of Atmospheric Pressure Plasma Jet}
\author{Nan Jiang, Ailing Ji and Zexian Cao}
\thanks{zxcao@aphy.iphy.ac.cn}
\affiliation{Beijing National Laboratory for Condensed Matters,
Institute of Physics, Chinese Academy of Sciences, Beijing 100190,
China}

\begin{abstract}
Atmospheric pressure plasma jet can protrude some 5.0 cm into air.
It holds promise for multivarious innovative applications, but its
formation mechanism remains unsettled. We show that the plasma jet
is essentially a streamer corona totally independent of, but
obscured by, dielectric barrier discharge. Consequently, the jets
can be equally successfully generated even with one single bare
metal electrode attached to the tube orifice, both downstream and
upstream simultaneously, and at a significantly reduced voltage.
These results will help understand the underlying physics and
facilitate a safer and more flexible implementation of this
marvelous plasma source.
\end{abstract}

 \maketitle


  The advantages of atmospheric pressure plasma over low pressure
discharges are well known. It can be dispensable with the expensive
vacuum operation and maintenance, thus it allows many innovative
designs to meet the growing demand for cost-effective, reliable and
easy-to-operate plasma sources \cite{re1,re2}. The atmospheric
pressure plasma jet (APPJ) is a newly invented, much valuable
non-thermal discharge that can protrude into the ambient air for
some 5.0 cm, and the ionic temperature is usually below $150
^{\circ}$C so that thermally sensitive materials can be treated.
These features imply a greatly enhanced applicability of APPJ in
materials processing, biomedicine, fabrication industries and so
forth \cite{re2,re3,re4}. Various APPJ devices, including jet needle
\cite{re5} and plasma pencil \cite{re6,re7,re8}, have been designed
since 2005. At the same time, the plasma jets are also found to
exhibit many intriguing characteristics.

 APPJ is an electrically driven phenomenon that merits in-depth research
in its own right. Recently, under a high-resolution intensified
charge coupled device the jet was found to fire smaller plasma
bullets at velocities in the order of $10^{4}\sim 10^{5} ~m/s$
\cite{re7,re8,re9,re10,re11,re12}. This triggers a new wave of
research enthusiasm towards APPJ
\cite{re5,re10,re11,re12,re13,re14}. Lu and Laroussi proposed a
model based on photoionization to explain the propagation kinetics
of the plasma bullets \cite{re7}, while Sands et al. recently
noticed that the plasma jet may be initiated independent of the
dielectric barrier discharge (DBD) and speculated that it is
streamer-like \cite{re12}. Yet the plasma jets are usually generated
by using the double electrode configuration for DBD, and the
mechanism for jet formation remains unsettled. In this letter we
show that to generate a plasma jet, the dielectric layer coating the
electrodes and the relevant DBD processes are totally irrelevant.
Rather, APPJ originates in a streamer corona. In recognition of this
fact, we succeeded in obtaining plasma jets with single dielectric
electrode or even single bare metal electrode attached to the tube
orifice, simultaneously in both downstream and upstream directions,
and, more importantly, at a significantly reduced voltage.

  Plasma jets concerned here were generated in a quartz capillary with an inner diameter of 2.0 mm and an outer
diameter of 3.5 mm. High-purity helium (5N) is used as the working
gas. A sinusoidal voltage at 17 kHz is applied for the excitation
and sustaining of the discharge. Three distinct electrode
configurations are applied to reveal the true origin of the plasma
jets: double dielectric electrodes (aluminum foil wrapping the
capillary), single dielectric electrode and single bare metal
electrode sitting at the tube orifice. Detailed geometry of the
electrodes will be specified at proper places. A digital camera
(Canon EOS 30D, with a 50 mm lens) and two photo-electron multiplier
tubes (PMT, Hamamatsu CR131) are used to study the optical emission
from the jets. The output of the PMTs is recorded by an oscilloscope
(Tektronix PDO 4032). The line along the two slits of PTM is
perpendicular to the quartz tube, and the ``vision" of the PMTs
spans only about 1.5 mm wide at the axis of the effluent. As having
been established by previous studies that the plasma jet comprises a
train of high-speed ``bullets", therefore when the plasma bullets
pass the front of the PMTs, the registered output will disclose the
form and propagation kinetics of the bullets, thus allowing  the
analysis of the origin for the plasma jet and the calculation of its
velocity.

\begin{figure}[t]
\includegraphics[width=7 cm,bb=-3 3 227 180]{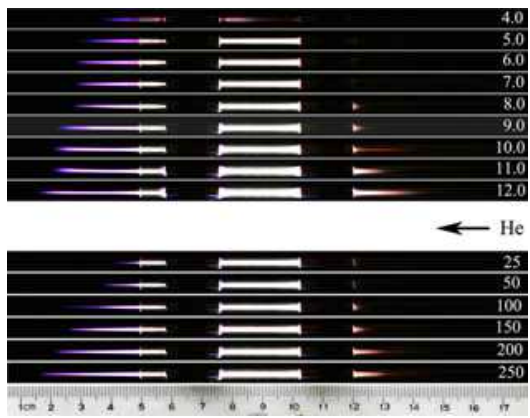}
\caption{Formation of discharge overfall as upstream plasma jet in
helium (right, pink colored) under the double electrode
configuration for DBD, with the ground electrode sitting on the
upstream side. Upper: at a constant flow rate of 150 liter/hour with
increasing applied voltage (in kV); Lower: at a constant voltage of
8.0 kV with increasing flow rate (in liter/hour). Electrode width:
1.8 cm; gap width: 3.0 cm.} \label{Fig1}
\end{figure}

  It begins with the observation of ``discharge overfall" accompanying the ignition of plasma jets with the DBD configuration (Fig.~\ref{Fig1}). Note in the upper
figure that when the voltage is below 4.0 kV (peak-to-peak value
throughout this letter), the discharge cannot fill the gap between
the electrodes, and the glow in the jet form appears symmetrically
on both flanks of the active electrode. In this case no discharge
current can be measured in the circuit connecting the two
electrodes. This strongly indicates that the discharge originates at
the active electrode. With increasing voltage, the jet grows in
length and the glow begins to fill the gap. Remarkably, for voltages
between $6.0\sim8.0$ kV, the jet length in air suffers only a
negligible change. It is also noticeable that from 8.0 kV on, a glow
in pink envelope appears in the overfall zone beyond the ground
electrode, and grows with increasing voltage.

  Now let's fix the voltage at 8.0 kV and vary the gas flow. Under given
conditions, the overfall always makes its presence and becomes
elongated with the increasing gas flow. The discharge overfall is
also a plasma jet; in fact, Lu and Laroussi used this effect to
generate downstream plasma jet, whereby the ground electrode sits
near the tube orifice, and the plasma plume also displays distinct
bullets \cite{re7}. The discharge overfall is obviously related to
the DBD processes, its occurrence here or its escaping others'
notice can be explained as follows. With a very large applied
voltage, the polarization charges at the ground electrode will get
saturated, and they are insufficient to compensate the discharge
process. Therefore the charge accumulation region will expand beyond
the ground electrode, leading to an overfall of charges. This is the
reason why a glow in the overfall region can be observable only when
the voltage exceeds 8.0 kV. The glow induced by charge overfall
becomes evidently elongated at larger voltages. As the amount of
polarization charges induced at electrode is proportional to the
electrode area, consequently a large-area ground electrode is
unfavorable for the formation of overfall. We noted that for a
ground electrode 5.0 cm wide, the threshold voltage needs be ~20 kV.
This explains why discharge overfall was not observed in previous
studies, e.g., in Ref. \cite{re9} where the ground electrode is 5.0
cm wide, whereas the voltage applied is below 15 kV.

\begin{figure}[t]
\includegraphics[width=7 cm]{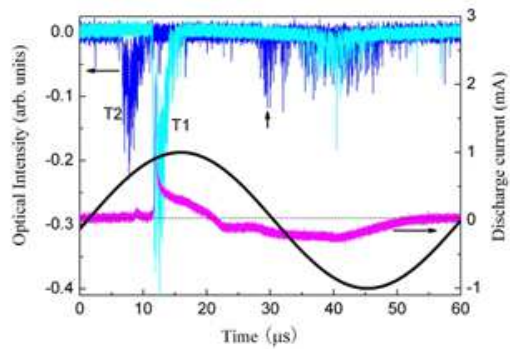}
\caption{Optical intensities monitored at 5 mm from the inner edge
of electrodes (T1 near ground electrode). Also the discharge current
inside one period of applied voltage is plotted. Arrow indicated a
second pulse occurred near active electrode. Applied voltage: 6.0
kV. Gas flow: 150 liter/hour.} \label{Fig2}
\end{figure}

 Measurement of the discharge current
and optical emission in between the electrodes tells more about the
nature of the discharge process. From Fig.~\ref{Fig2} we see that
the optical signal was first measured near the active electrode
which is currently positive in the first half period of the applied
voltage, and the discharge current peak lagged by 5.0 $\mu$s,
roughly in step with the large optical signal measured at the ground
electrode. This convinces us that the plasma jet propagates from the
transient anode to the cathode. The finely resolved optical signal
comprises sharp pricks whereas the discharge current varies in a
quite smooth manner. In the negative half period, the optical
signals measured near both electrodes are synchronous; they are now
much reduced in intensity and assume a dispersed distribution in
time, indicating a plasma column across the gap between electrodes.
Remarkably, in the first half period one more spark of minor
intensity was measured at the active electrode. In contrast,
identical optical emissions were measured on the two flanks of the
active electrode.

  The results presented in Figs.~1-2 concluded that the discharge starts
from the active electrode, and the discharge current keeps strict
pace with only the optical emission near the ground electrode. The
plasma jet thrusting into the air and that in between the electrodes
are essentially the same, whereas the overfall beyond the ground
electrode, whether upstream as here or downstream as in
Ref.~\cite{re7}, is of a different origin. These observations lead
to the speculation that the plasma jet originates in a streamer
corona instead of being a DBD in spite of the double electrode
configuration adopted, since by definition only a corona discharge
can be launched by one single electrode \cite{re15}. The plasma jet
is in fact corona streamer ignited by the strong field at the
neighborhood of active electrode, totally independent of  DBD
process. For DBD, the discharge forms a circuit with the two
electrodes, whereas for the jet, the circuit is formed via the
diffusion of the carriers in the gas to the virtual ground far away.
In the DBD setups previously studied, the DBD and the jet form their
own independent circuits. The nature of the latter is simply
screened and obscured. If streamer corona is the true responsible
mechanism, then APPJ must be equally launched by using single
electrode and further getting rid of the dielectric layer. In the
following we will demonstrate that by removing the ground electrode,
both downstream and upstream plasma jets can be launched
simultaneously under proper conditions. Moreover, a simple bare
metal electrode can do the job more efficiently.

\begin{figure}[t]
\includegraphics[width=7 cm]{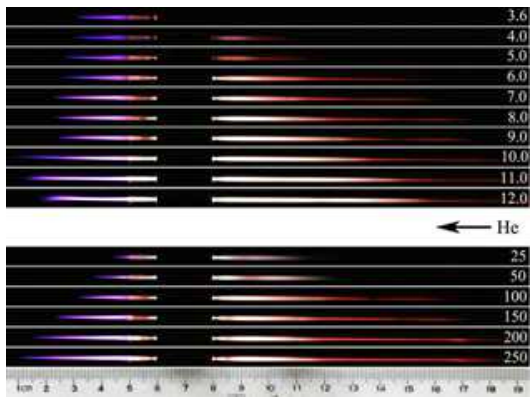}
\caption{Simultaneous ignition of downstream and upstream plasma
jets with single dielectric electrode.  Upper: flow rate fixed at
150 liter/hour; Lower: Voltage fixed at 8.0 kV. Electrode width: 2.0
cm.} \label{Fig3}
\end{figure}

  With single dielectric electrode,
plasma jets were simultaneously generated in both downstream and
upstream directions  under similar conditions (Fig.~\ref{Fig3}).
Naturally, now the DBD gap and the overfall region are no more
available. The downstream jet extends to a maximum 5.0 cm into the
air, whereas the upstream jet measures up to 12 cm. Roughly
speaking, the upstream jet assumes a much larger length but demands
a higher threshold voltage. The variation of jet length with the
applied voltage or the gas flow rate, and the differing behavior for
the upstream and downstream jets, substantiate the expectation that
the jet length hangs on the active region of the corona \cite{re15}.
Since inside the tube the disturbing air/helium interface is absent,
the relationship between the jet length and the applied voltage is
well expressed.

\begin{figure}[t]
\includegraphics[width=7 cm]{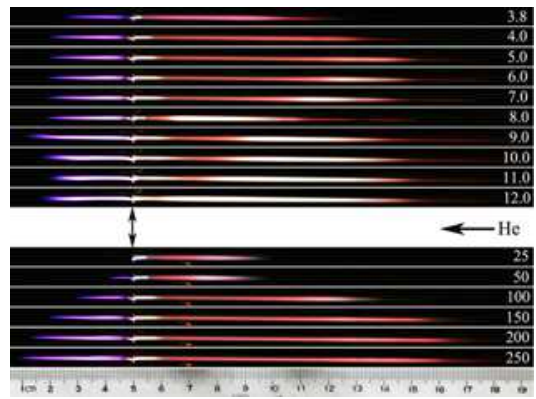}
\caption{Simultaneous ignition of downstream and upstream plasma
jets with single bare metal electrode attached to the tube orifice.
The electrode is made of a 0.5 mm thick metal foil with a hole of
2.5 mm in diameter. Upper: flow rate fixed at 150 liter/hour; Lower:
applied voltage fixed 5.0 kV.}
\label{Fig4}
\end{figure}

  Since it is the streamer mechanism that is responsible for the formation of APPJ, the dielectric barrier to the
electrode then bears no relevance. This is to say that a bare metal
electrode suffices for jet launching. Figure~\ref{Fig4} depicts the
APPJs generated with a metal electrode directly attached to the tube
orifice. As expected, both downstream and upstream jets were
successfully generated (Fig.~\ref{Fig4}), yet it show some dedicate
differences from the jets obtained with the previous two electrode
configurations. First, just with a voltage of 4 kV, the jet in the
ambient air approaches its maximum length limited by the air/helium
interface. This value is 9.0 kV for the double electrode
 and 7.0 kV for the single dielectric electrode
cases. The reason is that the direct contact between the electrode
and the discharge avoids the possible energy dissipation by the
dielectric layer. Second, at a fixed flow rate of 150 liter/hour,
the length of the upstream jet seems unperturbed for voltages over
5.0 kV, whereas in Fig.~3 the jet length steadily grows with the
voltage in the given range.

\begin{figure}[t]
\includegraphics[width=7 cm]{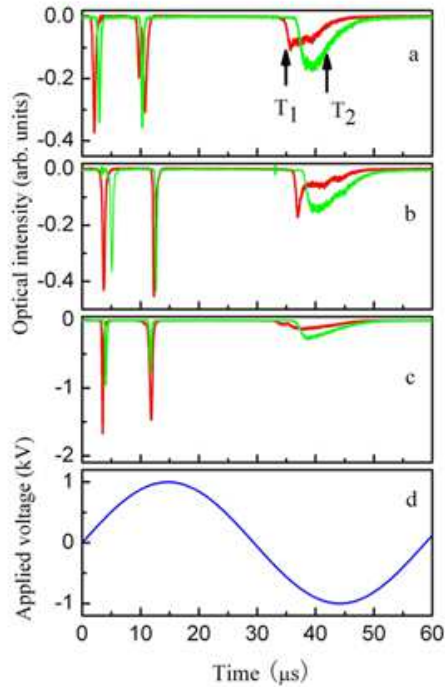}
\caption{Optical intensities (averaged over 64 periods) measured at
5 mm (T1) and 15 mm (T2) away from the tube orifice on the plasma
jets in air generated with (a) double-electrode configuration; (b)
single dielectric electrode; and (c) single bare metal electrode. In
(d) plotted is the applied voltage of 8.0 kV for reference. Flow
rate: 150 liter/hour. } \label{Fig5}
\end{figure}

  It would be of fundamental interest to compare the kinetic characters of the plasma jets
launched with the three different electrode configurations. The
temporal evolution of optical emission from the corresponding plasma
jets was illustrated in Fig.~\ref{Fig5}, which was measured at 5 mm
and 15 mm away from the orifice, respectively. We see that the three
curves display quite the same features. In the positive half-period,
two distinct pulses appeared. The presence of the second pulse is a
complicated story; it will be discussed elsewhere. A steady time
delay is established for the first peak, by which the jet velocity
can be determined. For the jets generated under the conditions given
in Fig.~5, the propagation velocity measures $1.1\times10^4$,
$0.7\times10^4$ and $1.9\times10^4~ m/s$, respectively. These values
are in agreement with Refs.~\cite{re9,re10}, whereas the larger
value in Refs.~\cite{re7,re12} in the order of $10^{5}~ m/s$ is due
to the application of a square-wave excitation (typical rise time at
20 ns) and a large overvoltage. From the optical emission
measurement and the true-color photographs of the jets, we concluded
that roughly the same plasma jets are obtained with the three
distinct electrode configurations.

  As confirmed above that the plasma jet is
in fact a streamer corona obscured by a capillary DBD, and as also
revealed by the direct observation of plasma bullets that the
temporal development and the structure of plasma jet source is
similar to that of a self-sustained streamer discharge in free space
\cite{re12}, the explanation of the jet velocity should be sought in
a model of streamer propagation. Dawson and Winn conceived the first
model for the cathode-directed streamer in 1965, and they obtained a
streamer propagation rate of $10^{5}$ $m/s$, but the head radius of
the streamer there is only $2.7\times10^{-5}$ $m$ \cite{re15,re16}.
Based on similar considerations, Lu proposed a modified version to
obtain a jet velocity as high as $10^{6}$ $m/s$ , but the radius of
the streamer head is required to be typically $7.0\times10^{-4}$ $m$
for the streamer to be able to self-propagate under low or zero
external field \cite{re7}.  In our case, the velocity is as low as
$1\sim2\times10^4$ $m/s$ due to the low and slow-varying voltage.
Clearly the non-uniformity of the electric field at the electrode
plays a pivotal role in determining the launching velocity of the
jet.  If the jet velocity is taken as the averaged velocity of
electron avalanches flying over a distance of the plasma sheath to
joint the ionic front, as given by $\mu$U/$\frac{}{}\chi$ , where U
is the potential of the streamer head, $\mu=0.113m^{2}/Vs$ is
electron mobility in helium, and $\chi$ measures the typical
distance of the origin of photoionized electrons to the center of
the spherical ionic front, then  assuming $\chi\sim3.0$ $mm$, this
gives a velocity of about $7.5\times10^{4}$ $m/s$ at $U=2$~kV.
(roughly corresponding to an applied voltage of 4.0 kV). This makes
a reasonable estimation to the measured jet velocities. Of course, a
more convincing model should include an exact knowledge of the
producing of avalanche on the photon path and of the avalanche
inception, thus to incorporate the indirect dependence on the shape
and the value of applied voltages.

 In summary, APPJs  generated in previous studies with double electrodes for DBD are essentially a corona
streamer. Consequently it would be more appropriately termed a
corona plasma jet. Plasma jets of comparable characteristics can be
effectively generated with all the three distinct electrode
configurations, and in both downstream and upstream directions
simultaneously. Particularly, the single bare metal electrode
permits a very flexible application of APPJ at reduced voltages. In
recognition of the streamer mechanism, a more reasonable qualitative
understanding of the plasma jet features has been obtained at the
moment; and it warrants a more fruitful exploration of the
underlying physics for APPJ which may exhibit  more interesting
properties. Other alternative methods for the generation of plasma
jets are to be expected.

This work was supported by NSFC grant no.10675163, and the National
Basic Research Program of China grant no.2009CB930800.

\end{document}